\documentclass[a4paper,12pt]{article}
\usepackage{amsmath,amssymb,epsfig}
\usepackage{graphics}
\usepackage{changebar}
\usepackage{bm}
\usepackage{graphicx}
\psfigdriver{dvips}

\def\bea{\begin{eqnarray}}
\def\eea{\end{eqnarray}}

\def\eq#1{Eq.~(\ref{#1})}

\def\Eqs#1#2{Eqs.~(\ref{#1}--\ref{#2})} 
\def\fig#1{Fig.~\ref{#1}}
\def\Fig#1{Fig.~\ref{#1}}

\def\jpar{j_{\parallel}}
\def\Bperp{\bm B_{\perp}}

\def\Vperp{\bm v_{\perp}}

\def\ez{\bm{e}_{z}}
\def\bnab{\bm{\nabla}}
\def\nabperp{\bnab}

\def\D'{\Delta'}
\def\leta{\ell_\eta}
\def\({\left(}
\def\){\right)}
\def\[{\left[}
\def\]{\right]}
\def\<{\left\langle}
\def\>{\right\rangle}

\def\d{\partial}

\begin{document}

\centerline {\footnotesize\em Proceedings of the 
31st EPS Conference on Plasma Physics, 
London 28 June -- 2 July 2004, 
Paper P-4.208}
\vskip0.5cm
\begin{center}
{\bf \large Fast and Slow Nonlinear Tearing Mode Reconnection}\\
\vskip0.2cm

N.~F.~Loureiro,$^a$  
S.~C.~Cowley,$^{a,b}$ W.~D.~Dorland,$^{c}$ \\
M.~G.~Haines$^a$ and A.~A.~Schekochihin$^d$
\vskip0.1cm

{\footnotesize 
\centerline{\em $^a$Department of Physics, Imperial College, London~SW7~2BW, UK}
\centerline{\em $^b$Department of Physics and Astronomy, UCLA, Los Angeles, CA 90024, USA}
\centerline{\em $^c$IREAP, University of Maryland, College Park, MD 20742-3511, USA}
\centerline{\em $^d$DAMTP, University of Cambridge, Cambridge CB3 0WA, UK}}
\end{center}

\vskip0.2cm

\noindent
{\bf 1. Introduction.} 
The standard theory of the tearing-mode evolution identifies three stages. 
The first is the linear stage described by the
Furth--Killeen--Rosenbluth (FKR) theory~\cite{FKR}. 
During this stage, the island width $W$ grows exponentially in time 
until it reaches the width of the resistive dissipation layer, 
$\leta\propto\eta^{2/5}\D'^{1/5}$, where $\eta$ is the resistivity
and $\D'$ is the instability parameter [see \eq{Dprime_def}]. 
Once $W\sim\leta$, nonlinear terms are sufficiently large to replace
inertia as the force opposing the inflow pattern. A slow down of the
growth ensues, from exponential to linear in time: 
$dW/dt\sim\eta\D'$. This is the second stage of the tearing-mode
evolution, known as the Rutherford regime~\cite{ruth_73}. 
Finally, the third, saturated, stage is reached 
when the island width becomes comparable to the equilibrium 
shear length~\cite{white_77}.

In this paper, we find 
the tearing-mode evolution to be, in fact, a four-stage 
process: the FKR regime, the Rutherford regime, 
a regime of fast nonlinear island growth that we 
identify as Sweet--Parker (SP) reconnection, and saturation. 
We carry out a set of numerical simulations 
that demonstrate two main points. 
First, we show that, given sufficiently small $\eta$,  
the Rutherford regime always exists; larger values of $\D'$ 
require smaller values of $\eta$. 
Rutherford's negligible-inertia assumption is validated and 
the asymptotically linear dependence of $dW/dt$ on $\eta$ 
and $\D'$ is confirmed. 
Second, we find that, at large $\D'$, 
the Rutherford regime is followed by a nonlinear stage of fast growth 
linked to $X$-point collapse and formation of a current sheet. 
This causes the reconnection to become SP-like. 
The signature $\eta^{1/2}$ scaling of the effective 
island growth rate is, indeed, found in this nonlinear stage. 
The SP stage culminates in the saturation of the mode, which can, 
thus, be achieved much faster than via Rutherford regime. 

\vskip0.2cm

\noindent
{\bf 2. The Model.} 
We use the conventional Reduced MHD set of equations~\cite{strauss_76} in 2D
for a plasma in the presence of a strong externally imposed magnetic 
field $B_z$:
\bea
\label{RMHD_vort}
\frac{\d \omega}{\d t} + \Vperp \cdot \nabperp \omega &=&
\Bperp \cdot \nabperp \jpar,\\
\label{RMHD_psi}
\frac{\d \psi}{\d t} + \Vperp \cdot \nabperp \psi &=&
\eta \nabperp^2 \psi,
\eea
where the total magnetic field is $\bm B = B_z\ez+\Bperp$, 
all gradients are in the $(x,y)$ plane, 
the in-plane magnetic feld is $\Bperp = \ez\times\bnab\psi$, 
the in-plane velocity is $\Vperp = \ez\times\bnab\phi$, 
and the parallel components of the vorticity and current are 
$\omega = \ez\cdot(\bnab\times\bm v) = \nabperp^2\phi$
and $\jpar= \ez\cdot(\bnab\times\bm B) = \nabperp^2\psi$. 
\Eqs{RMHD_vort}{RMHD_psi} are solved in a box with dimensions 
$L_x\times L_y$. 
All lengths can scaled so that the width of the box is $L_x=2\pi$. 

We impose an initial equilibrium defined by 
$\psi^{(0)}=\psi_{0}/\cosh^{2}(x)$ and $\phi^{(0)}=0$. 
We choose $\psi_0=1.3$ so that the maximum value of the 
unperturbed in-plane magnetic field $B_y^{(0)}=d\psi^{(0)}/dx$ 
is $1$. Time is scaled by the in-plane Alfv\'en time. 
The equilibrium is perturbed with 
$\psi^{(1)}=\psi_{1}(x)\cos(ky)$, where $k=mL_x/L_y$.
In our simulations, the initial perturbation has $m=1$. 
The island width is then approximately 
$W=4\sqrt{\psi_1(0)/\psi^{(0)\prime\prime}(0)}$. 
However, in what follows, $W$ is measured directly 
from the numerical data. 

With the equilibrium configuration we have chosen, 
the instability parameter is~\cite{porc_rec}:
\bea
\label{Dprime_def}
\Delta' = \frac{\psi'_1(+0)-\psi'_1(-0)}{\psi_1(0)} 
= 2\left[\frac{6k_{1}^{2}-9}{k_{1}(k_{1}^{2}-4)}-k_{1}\right],
\eea
where $k_{1}^{2}=k^{2}+4$. 
The equilibrium is tearing unstable if $k<\sqrt 5$. We vary 
the value of $\D'$ by adjusting the length of the box~$L_y$.  

\begin{figure}[t]
\unitlength1cm
\begin{minipage}[t]{8.0cm}
\epsfig{file=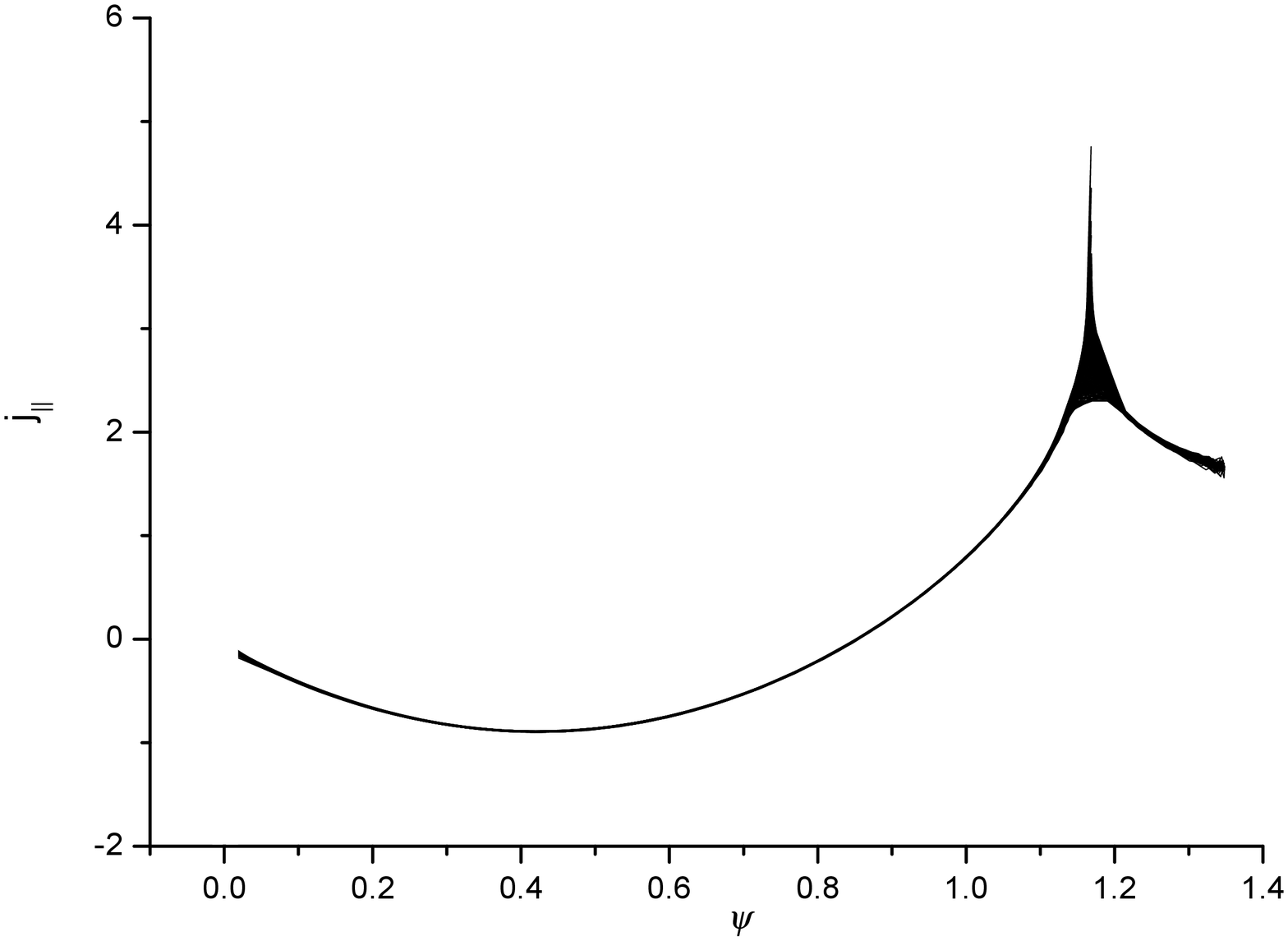,width=7.96cm}
\parbox[b]{7.5cm}{\caption{\label{j_psi} \footnotesize
Scatter plot of $\jpar$ vs.~$\psi$. Run parameters are $\D'=8.15, 
\eta=2.8\times10^{-4}$. Data extracted at $W=0.9$ [cf.~\fig{w_202}].
At the separatrix, $\psi\simeq1.17$.}}
\end{minipage}
\hfill
\begin{minipage}[t]{8.0cm}
\epsfig{file=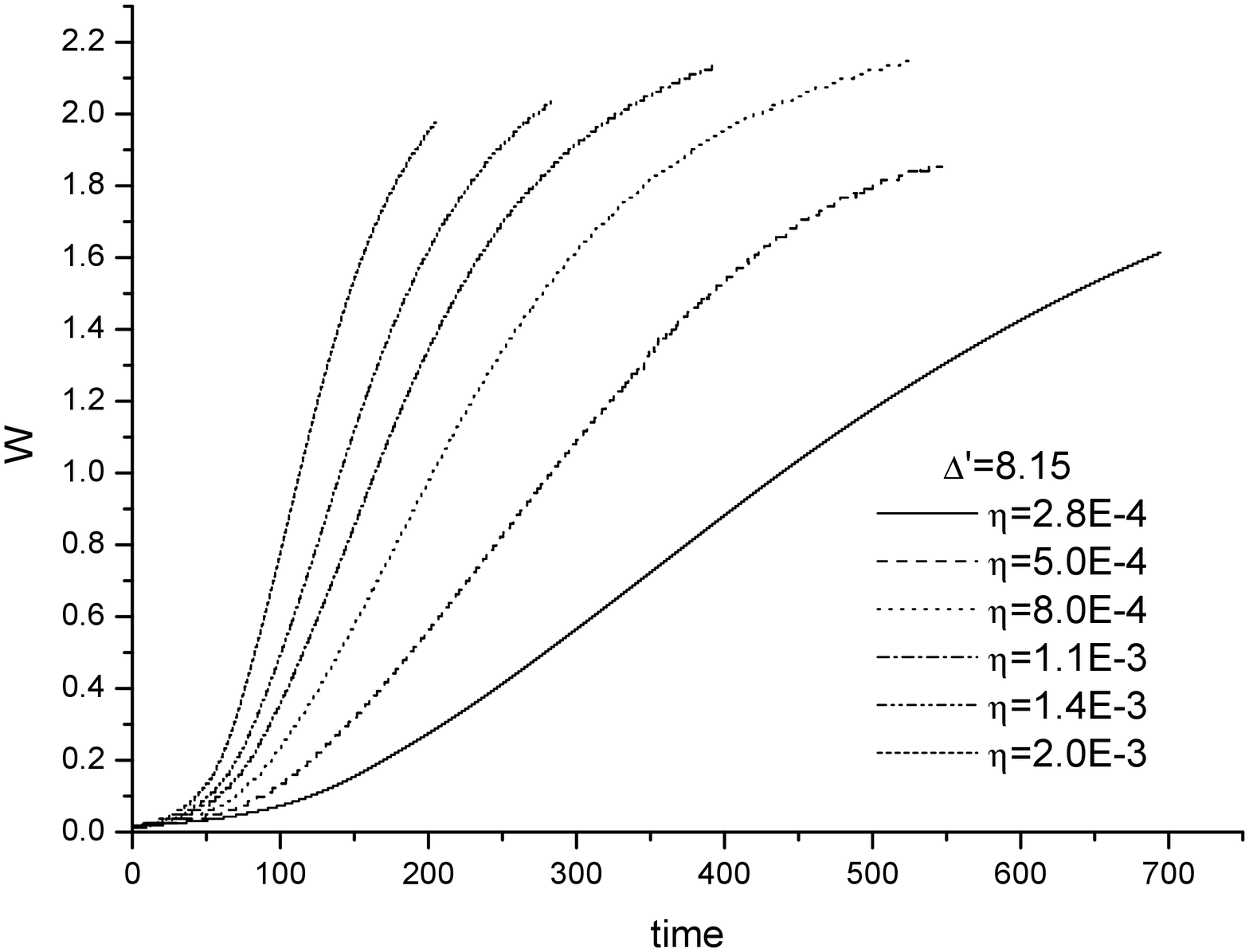,width=7.96cm}
\parbox[b]{7.5cm}{\caption{\label{w_202} \footnotesize
Island width vs.~time for $\D'=8.15$ and several values of $\eta$.}}
\end{minipage}
\end{figure}

\vskip0.2cm

\noindent 
{\bf 3. The Rutherford Regime.}
Rutherford's analysis depends on the assumption of negligible inertia, 
which reduces the vorticity equation to $\Bperp \cdot \bnab\jpar$=0. 
This implies that $\jpar=\jpar(\psi)$ everywhere except at the separatrix 
(the in-plane magnetic field vanishes at the $X$-point). 
\Fig{j_psi} proves the validity of this assumption. At the separatrix, 
$\psi\simeq1.17$. Larger values of $\psi$ correspond to the interior 
the island. The variation of $\psi$ in that region is relatively 
small, supporting the ``constant-$\psi$ approximation'' used by Rutherford.

\begin{figure}[t]
\unitlength1cm
\begin{minipage}[t]{8.0cm}
\epsfig{file=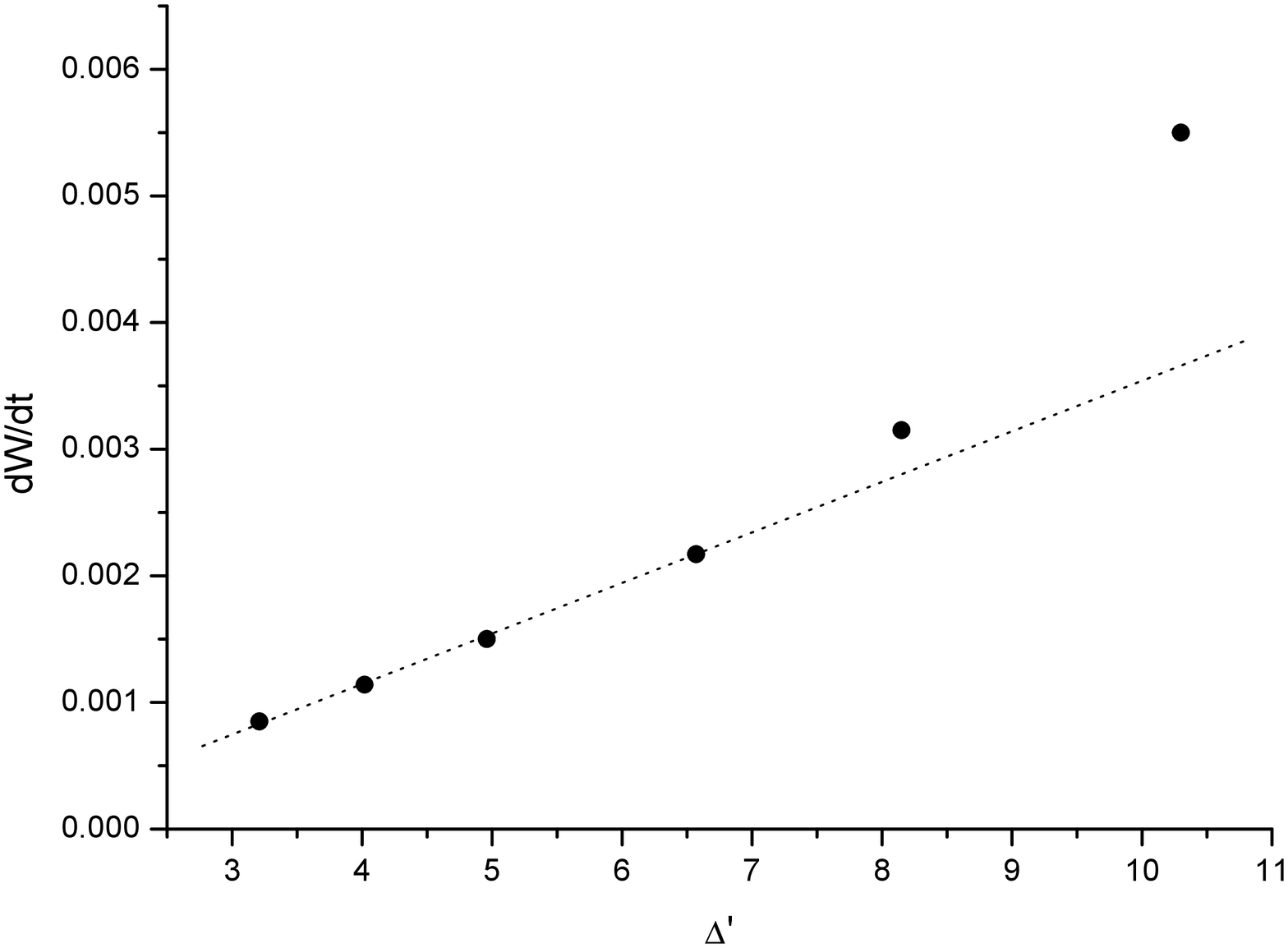,width=7.96cm}
\parbox[b]{7.5cm}{\caption{\label{delta_scaling} \footnotesize
$dW/dt$ vs.~$\D'$ at constant $\eta=2.8\times10^{-4}$.}}
\end{minipage}
\hfill
\begin{minipage}[t]{8.0cm}
\epsfig{file=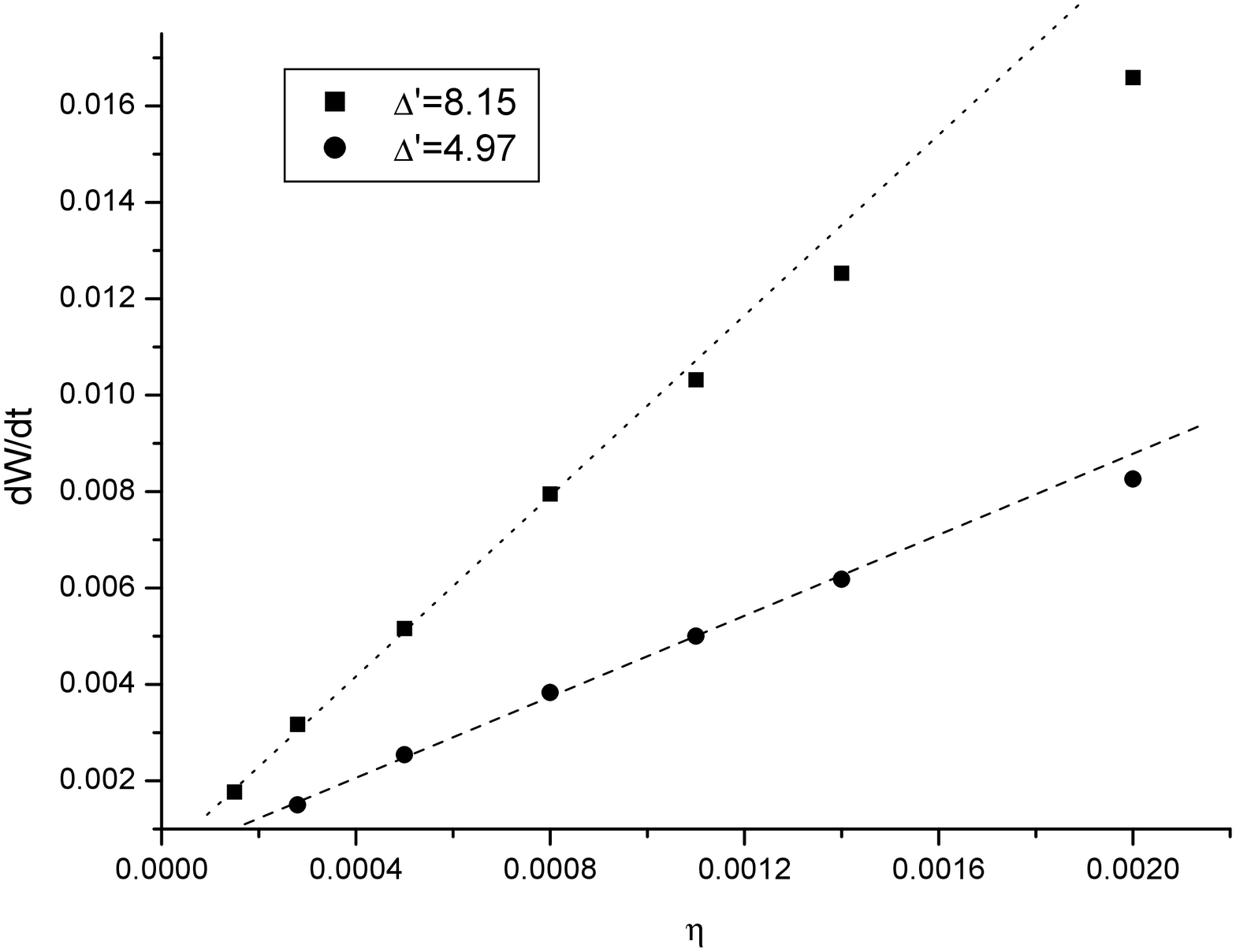,width=7.96cm}
\parbox[b]{7.5cm}{\caption{\label{eta_scaling} \footnotesize
$dW/dt$ vs.~$\eta$ at constant $\D'=4.97$, $8.15$.}}
\end{minipage}
\end{figure}

\Fig{w_202} shows the time evolution of the island width~$W$ at constant 
$\D'=8.15$ and varying resistivity. After the exponential growth stage, 
a distinct period of linear in time growth is manifest in all curves. 
Figs.~\ref{delta_scaling} and~\ref{eta_scaling} show 
the dependence of $dW/dt$ during the linear in time period 
on $\D'$ and $\eta$, respectively, 
demonstrating in both cases the linear relation predicted by Rutherford. 
At fixed finite $\eta$, the Rutherford scaling breaks down at large $\D'$. 
However, for a given $\D'$, it is recovered 
asymptotically at sufficiently small~$\eta$. 

Although Rutherford-like island growth 
has been observed in earlier numerical work, no parameter 
scan showing the linear scaling of $dW/dt$ with $\eta$ and $\D'$ has
previously been performed. White {\it et~al.}~\cite{white_IAEA} 
verified the linear in time growth of the island 
in their numerical simulations of the $m=2$ mode 
performed in tokamak geometry, 
a result later confirmed by Park {\it et~al.}~\cite{park_84}. 
Biskamp~\cite{bisk_NMHD} demonstrated the Rutherford behaviour in 
a numerical experiment done in slab geometry and with a
current-dependent resistivity. 
Biskamp's simulations had a relatively small value $\D'=3$.
Recently, Jemella {\it et~al.}~\cite{jemella} carried out a $\D'$ 
parameter scan with $\D'\in[0.92,20.93]$ (and constant $\eta$). 
Their results cast doubt upon the
validity of Rutherford's analysis by failing to produce the
linear in time behaviour: except for the smallest 
values of $\D'$, the island growth was exponential at all times. 
They argued that, instead of Rutherford's $X$-point configuration, 
a Sweet--Parker (SP) current sheet is formed and, consequently, 
SP reconnection replaces the Rutherford regime. 

We think that the absence of the Rutherford regime 
in Jemella {\em et al.}\ simulations is, in fact, 
due to the particular equilibrium configuration that they used: 
$B_y^{(0)}=\cos(x)$. 
We have also performed simulations with such a configuration and 
found the Rutherford stage absent, validating their results.  
The reason for this apparent 
sensitivity to the equilibrium profile 
is that a cosine configuration gives rise to two islands in the simulation 
box. In the nonlinear stage, circulating flows between the two $X$-points 
are established, which impedes the formation of well-separated nonlinear 
slow-down flows derived by Rutherford. 
We believe that, in order for the Rutherford regime to be recovered 
in the cosine configuration, even smaller values of $\eta$ 
than used by Jemella {\em et al.}\ (or by us) are necessary. 
Biskamp~\cite{bisk_NMHD}, who used $B_y^{(0)}=\tanh(x)$, 
did not have this problem and, accordingly, was able to see 
the Rutherford regime. 

\begin{figure}[h!]
\unitlength1cm
\rotatebox{90}{\includegraphics[width=5cm]{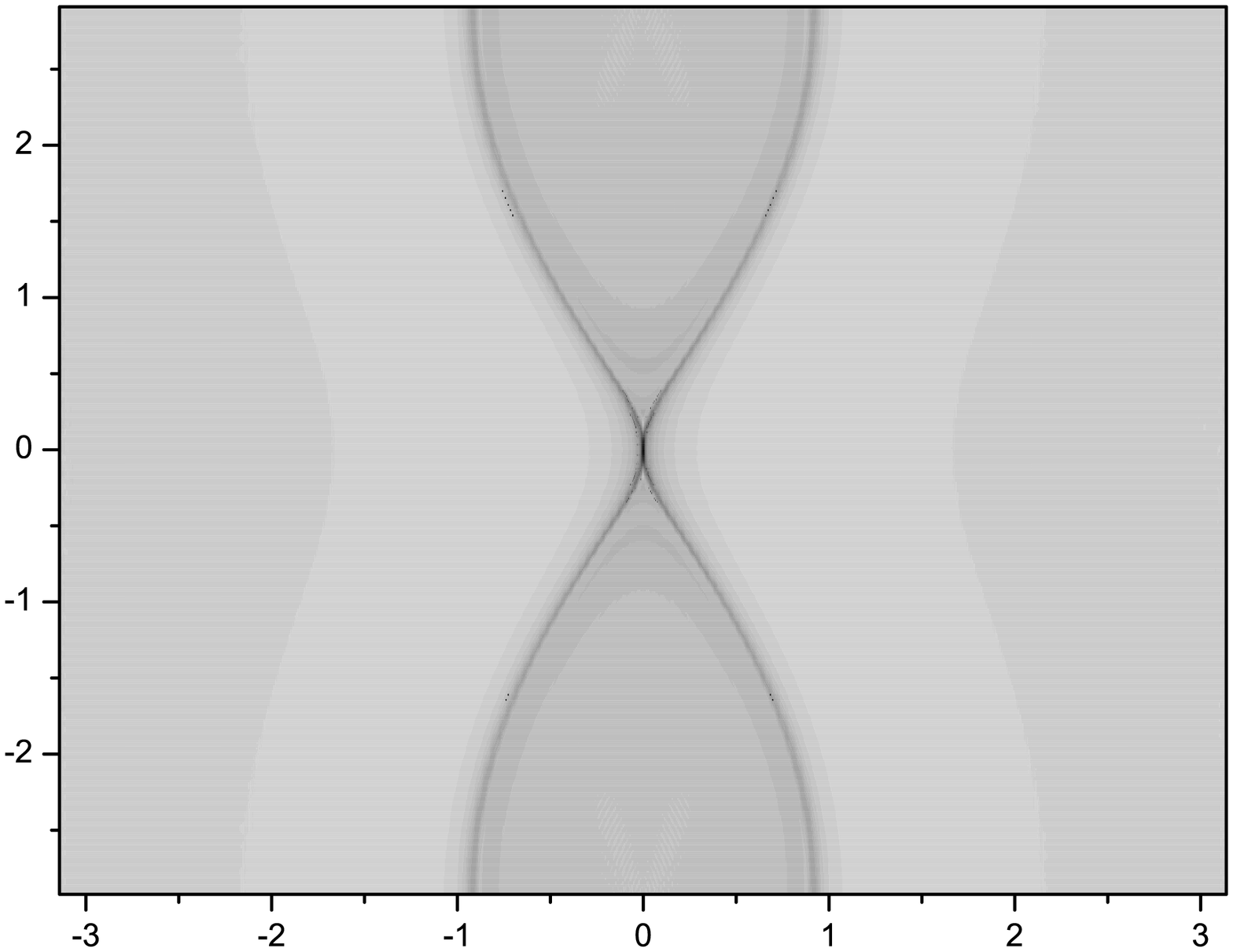}}
\hfill
\rotatebox{90}{\includegraphics[width=5cm]{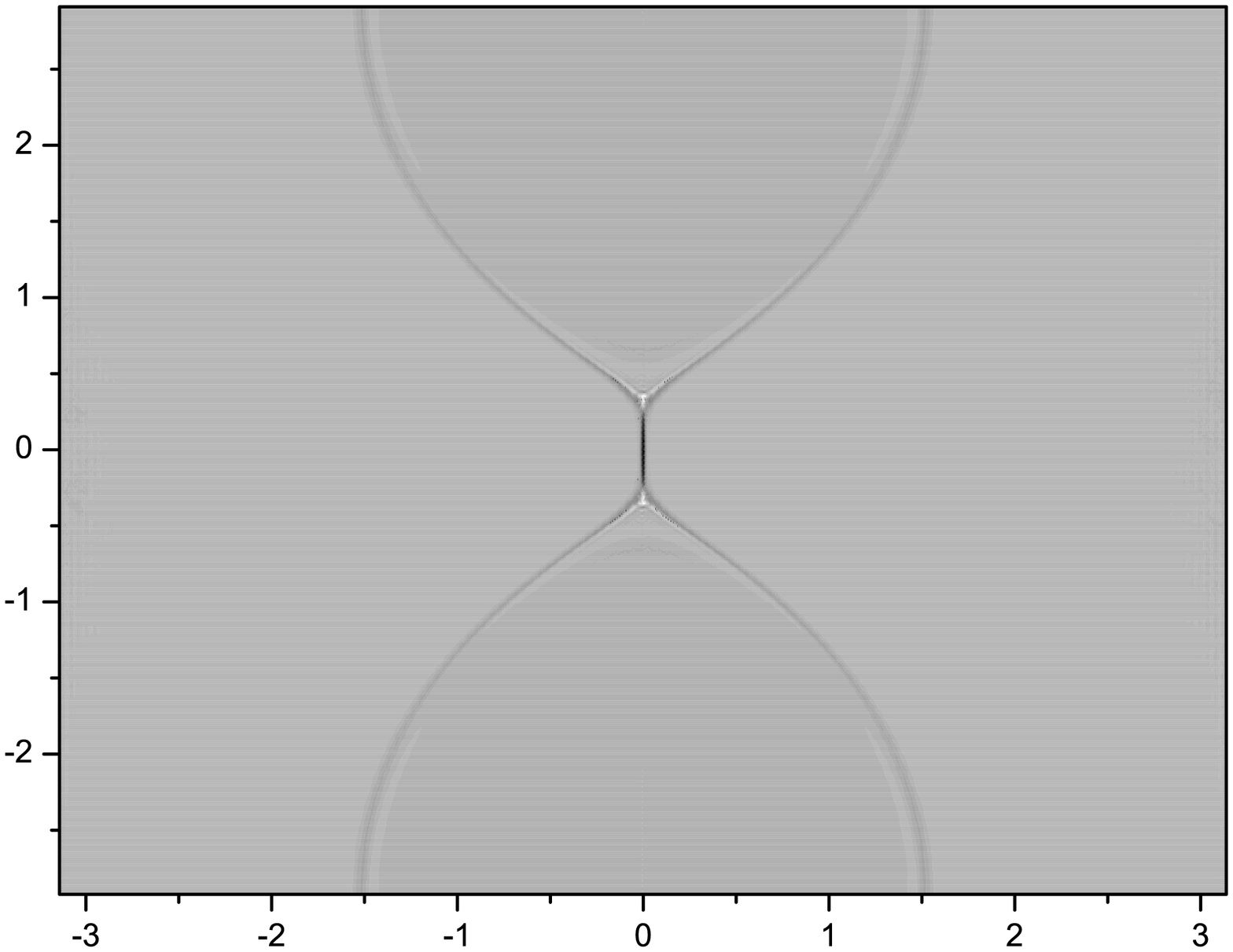}}
\caption{\label{contours} \footnotesize
Contour plots of the current $\jpar(x,y)$ 
at before (t=352, left panel) and after (t=385, right panel) 
for a run with $\D'=12.2$, $\eta=2.8\times10^{-4}$ 
the nonlinear speed up (the same run as in~\fig{growthrate_308}).} 
\end{figure}

\begin{figure}[t]
\unitlength1cm
\begin{minipage}[t]{8.0cm}
\epsfig{file=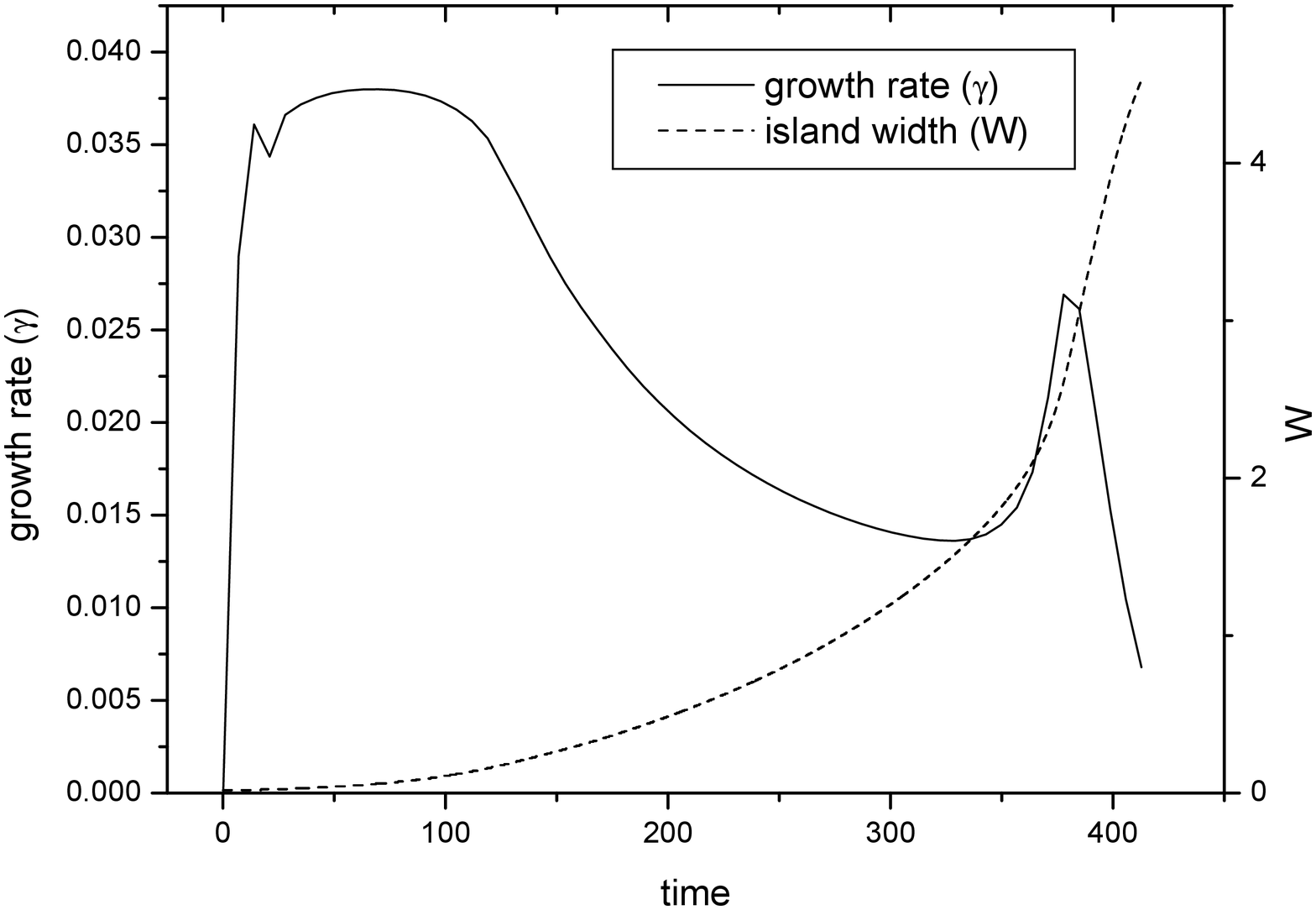,width=7.96cm}
\parbox[b]{7.5cm}{\caption{\label{growthrate_308} \footnotesize
Effective growth rate $\gamma$ 
and island width $W$ vs.~time, for 
a run with $\D'=12.2$ and $\eta=2.8\times10^{-4}$.}}
\end{minipage}
\hfill
\begin{minipage}[t]{8.0cm}
\epsfig{file=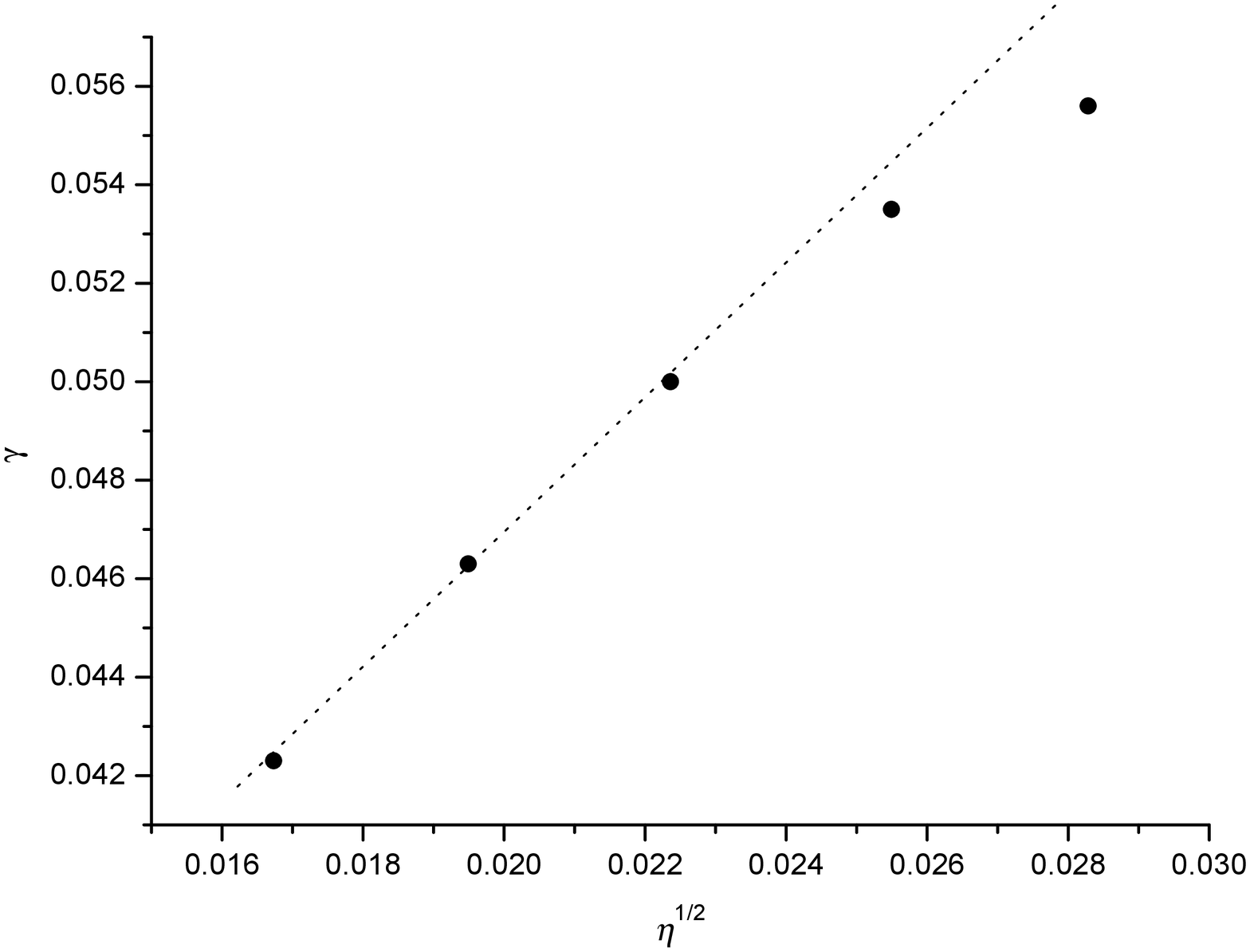,width=7.96cm}
\parbox[b]{7.5cm}{\caption{\label{SP_scaling} \footnotesize
Dependence of the peak effective growth rate $\gamma_{\rm max}$ 
in the fast nonlinear regime on $\eta^{1/2}$ 
at constant $\D'=17.3$.}}
\end{minipage}
\end{figure}

\noindent
{\bf 4. The Fast Nonlinear Stage.} 
In simulations with large $\D'$, we find that the $X$-point 
configuration maintained during the Rutherford regime 
eventually collapses and a current sheet is formed 
(\fig{contours}). A dramatic speed up of the island growth 
ensues (\fig{growthrate_308}). 
\Fig{SP_scaling} shows that the peak effective 
growth rate $\gamma=d\ln W/dt\sim 1/t$ 
during this fast nonlinear stage 
scales as $\eta^{1/2}$, thus supporting the expectation 
that the reconnection in this regime is of the SP kind.
This behaviour can be qualitatively understood by recalling 
that the $X$-point configuration assumed in 
Rutherford's analysis is, in fact, unstable~\cite{chap_63}. 
As time goes on, the effective island growth rate 
in the Rutherford regime decreases ($\gamma\sim 1/t$)
and eventually becomes so low that the $X$-point configuration 
can no longer be sustained over the time scale $\sim\gamma^{-1}$. 
$X$-point collapse leads to the formation of the current sheet. 

Note that, for large $\D'$, if $\eta$ is not sufficiently small for 
the Rutherford stage to occur, 
the SP stage nearly immediately follows the 
FKR stage, as is the case in Jemella {\em et al.}\ simulations. 
In the opposite case of small $\D'$ and small $\eta$, 
the saturation can be reached directly from the Rutherford 
stage, with the SP stage never having time to materialise. 

We leave a more detailed theoretical description of the 
nonlinear speed-up effect to a forthcoming paper. 

\vskip0.2cm

\noindent
{\footnotesize {\bf Acknowledgments.} 
Discussions with J.~F.~Drake, B.~N.~Rogers, and M.~A.~Shay are
gratefully aknowlegded. NFL was supported by Funda\c{c}\~ao para a Ci\^encia 
e a Tecnologia, Portuguese Ministry for Science and Higher Education. 
AAS was supported by the Leverhulme Trust 
via the UKAFF Fellowship.} 


{\footnotesize
\bibliography{lcdhs_EPS04}
\bibliographystyle{unsrt}}

\end{document}